\documentclass[amsmath,amssymb,aps,prl,reprint,floatfix]{revtex4-2}

\usepackage{amsmath}
\usepackage[usenames,dvipsnames]{xcolor}
\usepackage{mathtools}
\usepackage[colorlinks=true,citecolor=blue,linkcolor=blue]{hyperref}
\usepackage{bbm}
\usepackage{braket}
\usepackage{physics}
\usepackage{comment}
\usepackage{hyperref}
\usepackage{fancyref}
\usepackage{upgreek}
\usepackage{esint}
\usepackage{soul}
\usepackage{nicefrac}
\usepackage[normalem]{ ulem } 
\usepackage{floatrow}

\usepackage{graphicx}% Include figure files
\usepackage{dcolumn}% Align table columns on decimal point
\usepackage{bm}% bold math
\usepackage[none]{hyphenat}
\usepackage[mathlines]{lineno}% Enable numbering of text and display math
\usepackage [english]{babel}
\usepackage [autostyle, english = american]{csquotes}
\MakeOuterQuote{"}
\usepackage[nolist]{acronym}

\newcommand{\leri}[1]{\left(#1\right)}

\begin{document}

\title{The inverse Mpemba effect demonstrated on a single trapped ion qubit}
\author{Shahaf Aharony Shapira$^{\dagger}$}
\email{shahaf.aharony@weizmann.ac.il}
\author{Yotam Shapira$^{\dagger}$} 
\author{Jovan Markov}
\author{Gianluca Teza}
\author{Nitzan Akerman}
\author{Oren Raz}
\author{Roee Ozeri}
\affiliation{Department of Physics of Complex Systems\\
		Weizmann Institute of Science, Rehovot 7610001, Israel\\ 
		$^\dagger$ These authors contributed equally to this work
}

\begin{abstract}
The Mpemba effect is a counter-intuitive phenomena in which a hot system reaches a cold temperature faster than a colder system, under otherwise identical conditions. Here we propose a quantum analog of the Mpemba effect, on the simplest quantum system, a qubit. Specifically, we show it exhibits an inverse effect, in which a cold qubit reaches a hot temperature faster than a hot qubit. Furthermore, in our system a cold qubit can heat up exponentially faster, manifesting the strong version of the effect. This occurs only for sufficiently coherent systems, making this effect quantum mechanical, i.e. due to interference effects. We experimentally demonstrate our findings on a single $^{88}\text{Sr}^+$ trapped ion qubit. The existence of this anomalous relaxation effect in simple quantum systems reveals its fundamentality, and may have a role in designing and operating quantum information processing devices.
\end{abstract}

% \pacs{Valid PACS appear here}% PACS, the Physics and Astronomy
                             % Classification Scheme.
\maketitle
%---------------------------
\begin{acronym}
	\acro{ME}{Mpemba effect}
	\acro{GKSL}{Gorini-Kossakowski-Sudarshan-Lindblad}
	\acro{dof}{degree of freedom}
	\acro{ODE}{ordinary differential equation}
    \acro{QCs}{quantum computers}
    \acro{SM}{Supplemental Material}
\end{acronym}

Physical systems undergoing relaxation can exhibit a wide range of rich and non-trivial phenomena. A prominent example is the \ac{ME}~\cite{Mpemba1969Cool,jeng2006mpemba}, in which an initially hot system cools down faster than a colder, otherwise identical, system. Some systems manifest a stronger version of this effect~\cite{Klich2019Mpemba}, in which the hotter systems relaxes exponentially faster. The \ac{ME} has been experimentally demonstrated in various classical systems, e.g. water~\cite{jeng2006mpemba}, Clathrate hydrates~\cite{ahn2016experimental}, magnetic alloys~\cite{chaddah2010overtaking}, colloids diffusing in a potential~\cite{Kumar2020Exponentially} and a few others \cite{hu2018conformation,liu2023mpemba,chorazewski2023the}. An inverse-\ac{ME}, in which an initially colder system heats up faster than a warmer system, has been predicted~\cite{Lu2017Significance,Lasanta2017When} and recently measured~\cite{Kumar2022Anomalous}. Much theoretical insight was gained on this effect in recent years, using various theoretical methods \cite{Walker_2021,holtzman2022landau,degunther2022anomalous,busiello2021inducing,PhysRevLett.130.207103,schwarzendahl2022anomalous,zhang2022theoretical,teza2023relaxation,biswas2020mpemba,santos2020mpemba,chetrite2021metastable} and numerical results \cite{baity2019mpemba,gal2020precooling,gijon2019paths,vadakkayil2021should}.

Theoretical quantum versions of the \ac{ME} have been recently proposed in various models, e.g. Ising model~\cite{Nava2019Lindblad}, Anderson model~\cite{Chatterjee2023Quantum} and a perturbative technique for Markovian open systems~\cite{PhysRevE.108.014130}. Quantum Mpemba-like theories, which are non-thermal, have also been suggested, including accelerated relaxation of dissipative open systems~\cite{Carollo2021Exponentially,Kochsiek2022Accelerating} and relaxation of entanglement asymmetry in spin-systems~\cite{ares_entanglement_2023,Murciano2023Entanglement,rylands2023microscopic}. The latter has been recently demonstrated using trapped-ions~\cite{joshi2024observing}.

Here we propose and experimentally demonstrate the existence of an inverse-\ac{ME} in the simplest quantum system - a single qubit.  Our analysis shows that a strong inverse-\ac{ME} occurs for a sufficiently coherent qubit, making this effect quantum mechanical, i.e. due to interference.

We consider a coherently driven qubit that is coupled to a thermal Markovian bath, causing decoherence of the qubit and its eventual relaxation to a non-equilibrium steady state. Our only assumption on the qubit-bath coupling is that the qubit's decoherence rate is monotonically increasing with the bath's temperature. This occurs, e.g., for a black-body photon-emitting bath, such that the emission rate increases with temperature at every given wavelength, in particular at resonance with the qubit’s transition energy.

\begin{figure}[t]
    \includegraphics[width=\textwidth]{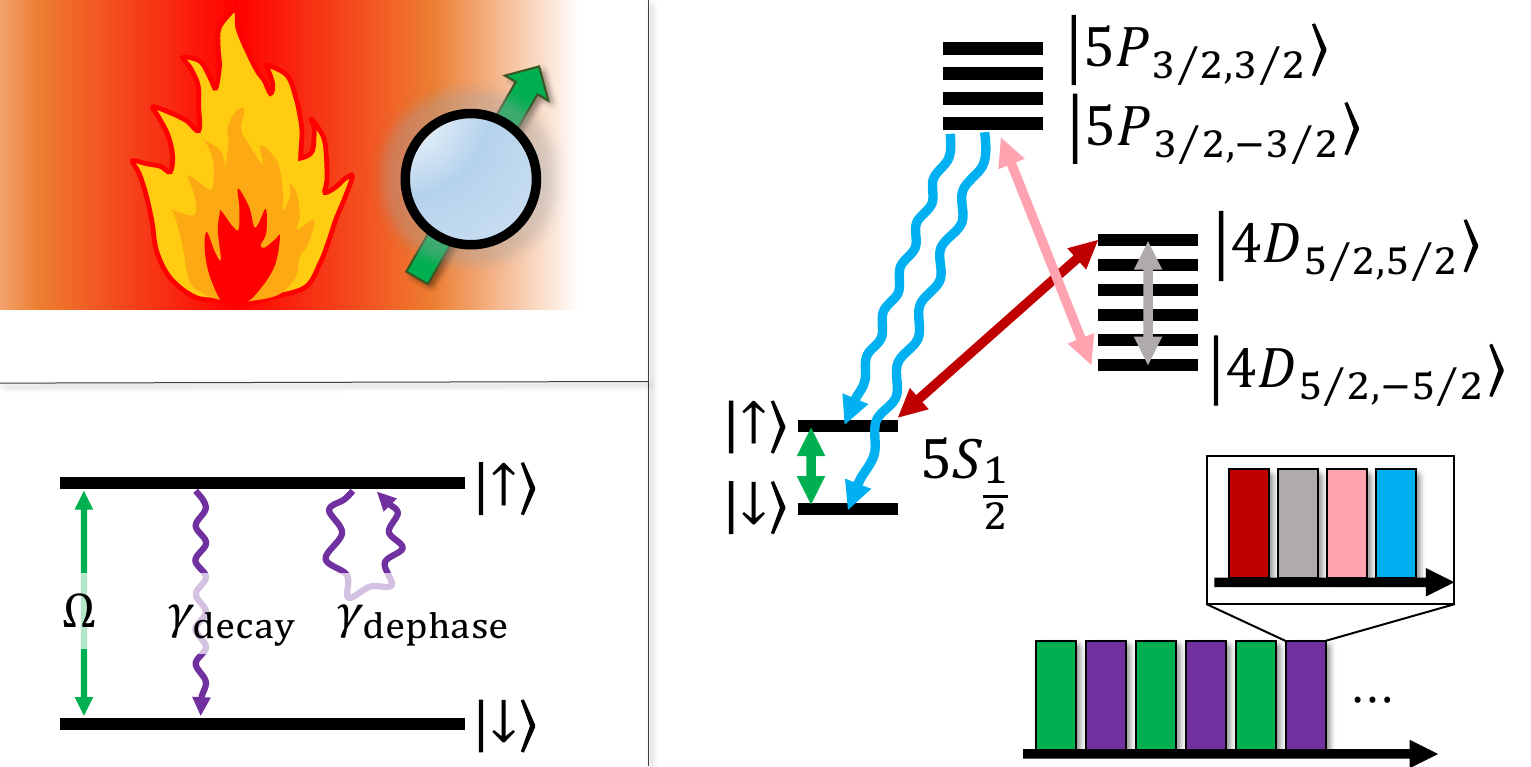}
    \caption{Top left: The modeled quantum system exhibiting an inverse Mpemba-effect. A thermal source of photons (fire) is coupled to coherently driven qubit (blue), causing it to relax to a steady state. Bottom left: The resulting coupling between the qubit's levels, $\ket{\downarrow}$ and $\ket{\uparrow}$, with a coherent drive (green) and decoherence terms causing decay ($\gamma_\text{decay}$) and dephasing ($\gamma_\text{dephase}$). Right: The qubit is mapped to the $5S_\frac{1}{2}$ levels of the Zeeman ground state manifold of a trapped $^{88}\text{Sr}^+$ ion. Coherent (green) and incoherent (purple) dynamics are interlaced in order to generate the overall system evolution (pulse sequence). The incoherent dynamics are generated using states in the long-lived $4D_\frac{5}{2}$ and short-lived $5P_\frac{3}{2}$ manifolds with various transitions (red, grey, pink and cyan), detailed below. 
    }
    \label{fig:Model}
\end{figure}

We demonstrate the inverse-\ac{ME} experimentally by implementing it on the Zeeman qubit, defined on a single trapped $^{88}\text{Sr}^+$ ion. Figure \ref{fig:Model} shows our model and the corresponding implementation on the ion's energy levels, detailed further below.

The qubit's dynamics is given by the \ac{GKSL} equation, $\partial_t\rho=\mathcal{L}\left[\rho\right]$, with $\mathcal{L}$ a Lindblad super-operator, acting on $\rho\in\mathbb{C}^{2\times2}$, the density matrix representing a statistical ensemble of a single qubit. Specifically, the operation of the super-operator is given by ($\hbar=1$),
\begin{align} \label{eqn:Lindbladian}
	\mathcal{L}\left[\rho\right] &= - \frac{i\Omega}{2} \comm{\sigma_x}{\rho} + \gamma_\text{decay}L_{\ketbra{\downarrow}{\uparrow}}\left[\rho\right]+ \gamma_\text{dephase}L_{\ketbra{\uparrow}{\uparrow}}\left[\rho\right]\,,
\end{align}
with $\Omega$ the rate of the coherent driving of the qubit, set by the $x$-Pauli matrix, $\sigma_x$. Open Markovian dynamics are generated by $L_A\left[\rho\right]\equiv A \rho A^\dagger -\frac{1}{2} \acomm{ A^\dagger A}{\rho}$. We consider decoherence due to decay (dephasing), generated by $\ketbra{\downarrow}{\uparrow}$ ($\ketbra{\uparrow}{\uparrow}$), with rates,
\begin{align}
	\gamma_\text{decay} &= \alpha \gamma\leri{T} \,, &
	\gamma_\text{dephase} &= \leri{1-\alpha} \gamma\leri{T} \,,
\end{align}
where $\alpha \in [0,1]$ is the relative occurrence of decay with respect to the decoherence rate and $\gamma\leri{T}$ is the overall temperature-dependent decoherence rate due to coupling to the bath~\cite{McCauley2020}. Throughout this work, the qubit's temperature is defined once it reaches a steady state with the bath. We note that $\gamma(T)$ is assumed to be monotonically increasing with $T$, e.g. via Planck's law, such that we can characterize the bath and the steady state temperature by $T$ or $\gamma$ interchangeably.

\begin{figure}[t]
    \centering
    \includegraphics[width=1\linewidth]{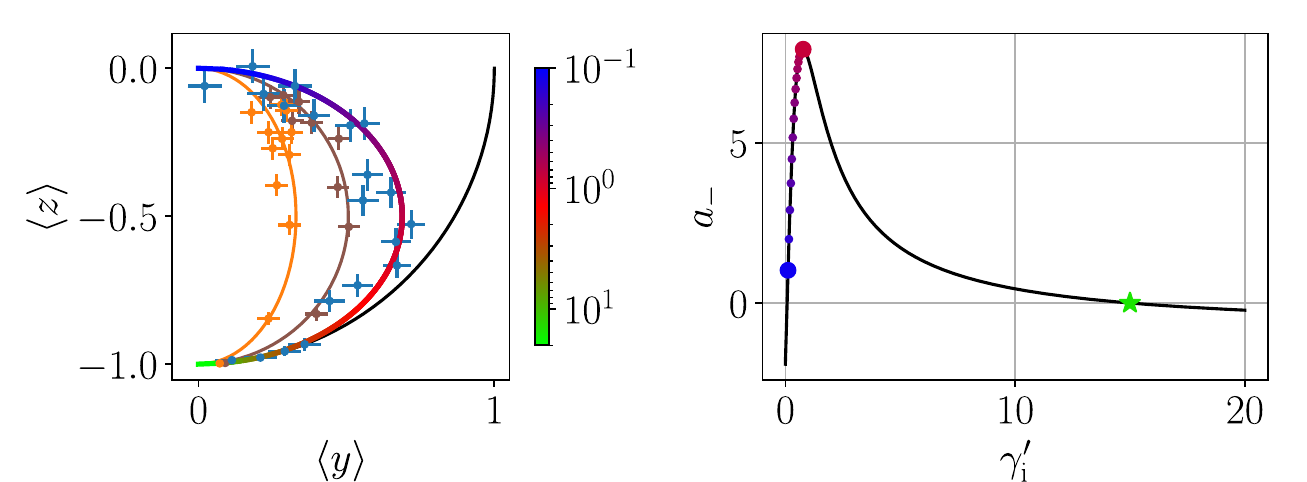}
    \caption{Left: Steady state locus. We measure the qubit's steady-state position (points) on the y-z plane of the Bloch sphere (black line) at different temperatures and compute the corresponding $\alpha$'s (see the main text). Our data-sets (points) are fitted yielding  $\alpha=0.21\pm0.03$ (orange), $\alpha=0.51\pm0.04$ (brown) and $\alpha=0.94\pm0.07$ (blue), used hereafter. The steady state locus corresponding to the latter is presented in color, showing values of $\gamma^\prime$ (log-scale). Error bars correspond to $\pm 2\sigma$ confidence intervals due to the quantum shot-noise of 300 experimental repetitions. Right: Coefficient of the slow-decaying eigenstate, $a_-$, as a function of $\gamma_\text{i}^\prime$, for $\gamma_\text{f}'=15$ (green star). The coefficient shows a non-monotonic behaviour, implying the existence of a \ac{ME}. Furthermore, the curve is shown to vanish at $\gamma_\text{i}^\prime = 0.07$, proving the existence of a strong-\ac{ME}. The highlighted points correspond to the $\gamma_\text{i}^\prime$s in Fig. \ref{fig:DynamicsDigital}.}
    \label{fig:SphereAminus}
\end{figure}

The dynamics is conveniently analyzed using the Bloch vector, $\vec{r}=\leri{x,y,z}$, with $\rho=\nicefrac{1}{2}\leri{1+\vec{r}\cdot\vec{\sigma}}$ (See derivation in the SM \cite{SM}). Since the system is driven, its fixed points correspond to non-equilibrium steady states that do not obey detailed balance, e.g. the qubit continuously scatters photons and its long-time limit is not $\propto e^{-\beta H}$. The collection of steady-states, $\vec{r}^\text{ ss}\leri{\gamma}$, form a right-half of an ellipse in the $y-z$ plane, with its center at $\leri{0, 0,-\nicefrac{1}{2}}$ and semiaxes $\leri{r_y,r_z}=\leri{\sqrt{\nicefrac{\alpha}{2}}, \nicefrac{1}{2}}$, shown in Fig.~\ref{fig:SphereAminus} (left). Each point on this curve, known as the steady state locus, corresponds to a steady state at a given $\gamma^\prime\equiv\gamma/\Omega$, with $\gamma^\prime\to0\leri{\infty}$ corresponding to the center (south-pole) of the Bloch sphere.

Consider the relaxation path of an initial condition given by the steady state solution of a cold temperature, $\vec{r}^\text{ ss}\leri{\gamma_\text{i}}$, when coupled to a hot bath characterized by $\gamma_\text{f}$. The solution of Eq.~\eqref{eqn:Lindbladian} is then given by,
\begin{equation}
    \vec{r}\leri{t;\gamma_\text{i},\gamma_\text{f}}=\vec{r}^\text{ ss}\leri{\gamma_\text{f}}+\sum_{n\in\left\{+,-,x\right\}} a_n\leri{\gamma_\text{i},\gamma_\text{f}} \vec{v}_n\leri{\gamma_\text{f}}e^{\lambda_n\leri{\gamma_\text{f}}t},\label{eqn:Solution}
\end{equation}
where $\vec{v}_n\leri{\gamma_\text{f}}$ are the relaxation modes of the system, $\lambda_n\leri{\gamma_\text{f}}$ their rates, and $a_n\leri{\gamma_\text{i},\gamma_\text{f}}$ the corresponding coefficients, determined by the overlap between the initial state and $\vec{v}_n(\gamma_\text{f})$. 

We note that the $x$-coordinate has a stable fixed point at $x^*=0$, making the $x$-direction trivially vanish throughout the system's evolution.

The decay rates in the $y-z$ plane are given by the real part $\Re\left[ \lambda_\pm \right]$, with
\begin{equation}
    \lambda_{\pm} = -\gamma_\text{f} \leri{\alpha + \nicefrac{1}{2} \pm \sqrt{\leri{\alpha - \nicefrac{1}{2}}^2 - \nicefrac{1}{\gamma_\text{f}^{'2}}}}\label{eqnEignevalues}.
\end{equation}
The \ac{ME} can exist only when $\Re\left[ \lambda_\pm \right]$ are distinct, allowing for a slow and fast relaxation modes. This occurs for final temperatures $\gamma_\text{f}^\prime>\gamma_\text{b}^\prime$, with the bifurcation point $\gamma_\text{b}^\prime \equiv \left| \alpha - \nicefrac{1}{2}\right|^{-1}$.

The relaxation at long times is determined by the slowest relaxation mode, $\lambda_-$, and its coefficient, $a_-$, which clearly vanishes for $\gamma_\text{i} = \gamma_\text{f}$. Fixing $\gamma_\text{f}^\prime$, one might expect $a_-$ to be monotonic in the range, $0\leq\gamma_\text{i}^\prime\leq\gamma_\text{f}^\prime$. However, for an inverse-\ac{ME} to take place, a cold system must reach the steady state faster than a hotter one, i.e. $\left| a_- \right|$ is smaller for a cold system, compared to a hotter system. It is therefore the non-monotonic behavior of $a_-$ as a function of $\gamma_\text{i}$ which enables the existence of the \ac{ME}~\cite{Klich2019Mpemba}. 
Indeed, the coefficient $a_-\leri{\gamma_\text{i},\gamma_\text{f}}$ displays such a behavior, implying the existence of an inverse-\ac{ME}. An example with $\gamma_\text{f}^\prime=15$ is plotted in Fig. \ref{fig:SphereAminus} (right). 

A strong-\ac{ME} occurs in the special case in which $a_-$ vanishes at an initial temperature, $\gamma_\text{i,SME}\neq\gamma_\text{f}$. In that case, the relaxation time is determined by the fast rate, $\lambda_+$, and as a result, it is exponentially faster~\cite{Klich2019Mpemba}. In other words, defining the distance to steady state,  $d_\text{ss}^{\gamma_\text{i}}\leri{t}\equiv\left|\vec{r}\leri{t;\gamma_\text{i},\gamma_\text{f}}-\vec{r}^\text{ ss}\leri{\gamma_\text{f}}\right|$, then $d_\text{ss}^{\gamma_\text{i}\neq\gamma_\text{i,SME}}\leri{t}/d_\text{ss}^{\gamma_\text{i,SME}}\leri{t}$ is asymptotically exponentially increasing in time.

Here, $a_-$ vanishes at $\gamma_\text{i,SME}^\prime = \gamma_\text{f}^\prime \leri{\leri{\alpha-\nicefrac{1}{2}} - \sqrt{\leri{\alpha-\nicefrac{1}{2}}^2 - \gamma_f^{'-2}}}$. For example, for $\gamma_\text{f}^\prime=15$, $a_-$ vanishes at $\gamma_\text{i,SME}'\approx0.07$ as seen in Fig.~\ref{fig:SphereAminus} (right). The strong-\ac{ME} in this system is experimentally optimal to achieve the most pronounced signal. Indeed, in our experimental demonstrations we make use of values $\gamma_\text{i}\approx\gamma_\text{i,SME}$. We mathematically prove the strong-\ac{ME} can only appear in a heating process, i.e. as an inverse-\ac{ME} (see the SM~\cite{SM}).

Since $\gamma_\text{i,SME}^\prime>0$, the required $\alpha$ for a strong-\ac{ME} is bounded by $\alpha > \nicefrac{1}{2}+ \nicefrac{1}{\gamma_\text{f}'} \geq \nicefrac{1}{2}$. When $\alpha$ satisfies this condition, there exists a strong-\ac{ME} for every final temperature above the bifurcation point, at $\gamma_\text{i,SME}'$. Thus, an exponentially faster relaxation occurs only in a sufficiently coherent system, i.e. with a low excess dephasing on top of that induced by the decay channel. Specifically, a classical bit, with no coherence between its two states, cannot exhibit this strong effect.

This model describes, for example, a single trapped $^{88}\text{Sr}^+$ ion qubit in a small-scale quantum computer~\cite{Manovitz2022Trapped}. Specifically we encode the $\ket{\downarrow}$ ($\ket{\uparrow}$) qubit state on the $5S_{\frac{1}{2},-\frac{1}{2}}$ ($5S_{\frac{1}{2},\frac{1}{2}}$) state in the Zeeman ground state manifold, shown in Fig.~\ref{fig:Model} (right). The two states are coherently coupled with a magnetic field (green), oscillating at the Zeeman splitting frequency in the $5S_{\frac{1}{2}}$ manifold, generating the qubit's Hamiltonian, $H=\Omega\sigma_x$, with $\Omega$ the field's Rabi frequency. 

As shown in Fig.~\ref{fig:Model}, we combine the coherent and open dynamics in discrete steps, by interlacing small durations of coherent (green pulse) and open Markovian evolution (purple pulse), i.e. by trotterization. Markovian open dynamics are generated by coupling the qubit levels via fast decaying states ~\cite{BenAv2020Direct}. Control over $\gamma$ and $\alpha$ is gained by making use of sequential cascade of pulses and transitions. 

Specifically we use a narrow linewidth laser at 674 nm ~\cite{Peleg2019phase} (red) in order to selectively couple the $\ket{\uparrow}$ state to the $\ket{4D_{\frac{5}{2},\frac{5}{2}}}$ state in the $4D_\frac{5}{2}$ metastable manifold. An additional laser at 1033 nm (pink) couples the $4D_\frac{5}{2}$ manifold to the short-lived $5P_\frac{3}{2}$. Due to selection rules, only the $\ket{5P_{\frac{3}{2},\frac{3}{2}}}$ state is populated, which quickly decays back to the $\ket{\uparrow}$ state (blue), resulting in full dephasing, i.e. $\alpha=0$. By using an additional  $\pi$-pulse in the $4D_\frac{5}{2}$ manifold (grey), between the $674$ nm and the $1033$ nm pulses, we map the $\ket{4D_{\frac{5}{2},\frac{5}{2}}}$ state to the  $\ket{4D_{\frac{5}{2},-\frac{5}{2}}}$, which will ultimately decay to the $\ket{\downarrow}$ state, yielding $\alpha\approx1$. The value of $\gamma$ is determined by the 674 nm pulse amplitude and length, as these control the relative population that is excited outside of the $5S_\frac{1}{2}$ qubit manifold in each pulse cycle.

We demonstrate this control experimentally by initializing the system to the $\ket{\uparrow}$ state and letting it relax to a steady state under $n=100$ repetitions of interlaced dynamics, analogous to a decay time of $7\gamma_\text{f}^{-1}$. After this evolution we perform state tomography to determine the location of the steady state on the Bloch sphere. Figure \ref{fig:SphereAminus} (left) shows the measured steady states for various values of $\gamma^\prime$, forming the elliptically shaped steady state locus (blue points), with a fitted value of $\alpha=0.94\pm0.07$ (gradient line).   

Intermediate values of $\alpha$ can be formed by replacing the $\pi$-pulse in the $4D_\frac{5}{2}$ manifold (grey) with, e.g., a $\pi/2$-pulse or a $\pi/5$-pulse, yielding a thinner steady state loci, fitted as $\alpha=0.51\pm 0.04$ and $\alpha=0.21\pm0.03$, (brown and orange) respectively.

A canonical experimental protocol for measuring the \ac{ME}, comprises letting the system relax to the steady state $\vec{r}^\text{ ss}\leri{\gamma_\text{i}^\prime}$, then quenching it to a final temperature, $\gamma_\text{f}^\prime$, while performing tomography of the relaxation dynamics to $\vec{r}^\text{ ss}\leri{\gamma_\text{f}^\prime}$. The measurements are then used to obtain the Euclidean-distance on the Bloch sphere from the final steady state, $d_\text{ss}\leri{t}$. This protocol raises a technical challenge, namely, the relaxation time to an initial cold system, with $\gamma_\text{i}^\prime\ll1$, requires a long evolution duration, which may surpass the system's natural coherence time, leading to an effectively reduced and uncontrolled value of $\alpha$.

We mitigate this challenge by measuring the \ac{ME} using two complementary techniques: performing the experimental protocol with large trotter steps, thus reducing the total duration of an experiment, or by effectively preparing the qubit in the initial state, $\vec{r}^\text{ ss}\leri{\gamma_\text{i}^\prime}$, thus circumventing the long initial relaxation time.

\begin{figure}[t]
    \includegraphics[width=\textwidth]{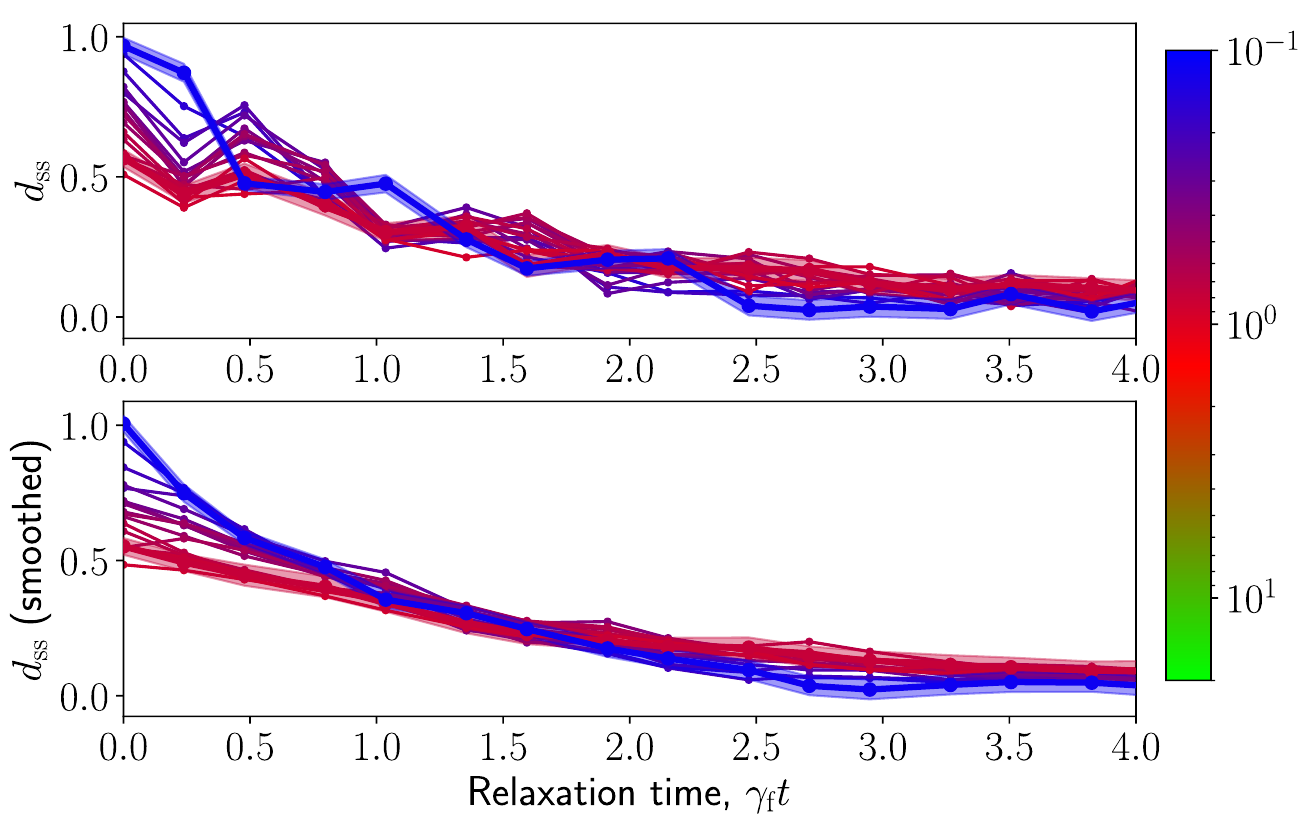}   
    \caption{The inverse-\ac{ME} is demonstrated by relaxing qubits to a steady state at various initial temperatures, $\gamma_\text{i}^\prime$ (color), and tracking their relaxation as a function of time (horizontal) to a final steady state at a fixed temperature, $\gamma_\text{f}^\prime=15>\gamma_\text{i}^\prime$ and $\alpha=0.94$. We consider the qubit's Euclidean-distance to the final steady state, $d_\text{ss}\leri{t}$ (vertical). We highlight an initially cold (thick blue) and hot (thick red) systems, $d_\text{ss}^\text{C}$ and $d_\text{ss}^{\text{H}}$, analyzed further below. Error regions correspond to $\pm 2\sigma$ confidence intervals due to the quantum shot-noise of 400 experimental repetitions. Top: $d_\text{ss}$  exhibits oscillations, not captured by the model above, which occur due to the relatively large time steps used in the evolution. Bottom: Post-processing the same data by polynomial smoothing reproduces the inverse-\ac{ME}. Specifically, $d_\text{ss}^{C}>d_\text{ss}^{H}$ at $t=0$, however their values cross at $t\approx2\gamma_\text{f}^{-1}$, after which $d_\text{ss}^{C}<d_\text{ss}^{H}$.}
    \label{fig:DynamicsDigital}
\end{figure} 

The results obtained by the former technique, large trotter steps, are shown in Fig.~\ref{fig:DynamicsDigital}. Specifically we evolve the system to $t=4\gamma_\text{f}^{-1}$ in 14 trotter-steps (horizontal) and present $d_\text{ss}\leri{t}$ (vertical) for various $\gamma_\text{i}^\prime$'s (color). These steps form a less accurate approximation of the continuous model. Indeed, Fig.~\ref{fig:DynamicsDigital} (top) exhibits oscillations which are common to non-adiabatic digitized evolution. We compensate for the oscillations by employing simple polynomial-smoothing of the data, shown in Fig.~\ref{fig:DynamicsDigital} (bottom). 

We highlight an initially cold system, $\gamma_\text{i}^{\prime \text{C}}=0.116$ (blue) and an initially hot system, $\gamma_\text{i}^{\prime \text{H}}=0.776$ (red). These demonstrate an inverse-\ac{ME}, as the curves of the cold and hot systems cross, with the cold system reaching the steady state before the hot system. 

\begin{figure}[t]
    \includegraphics[width=\textwidth]{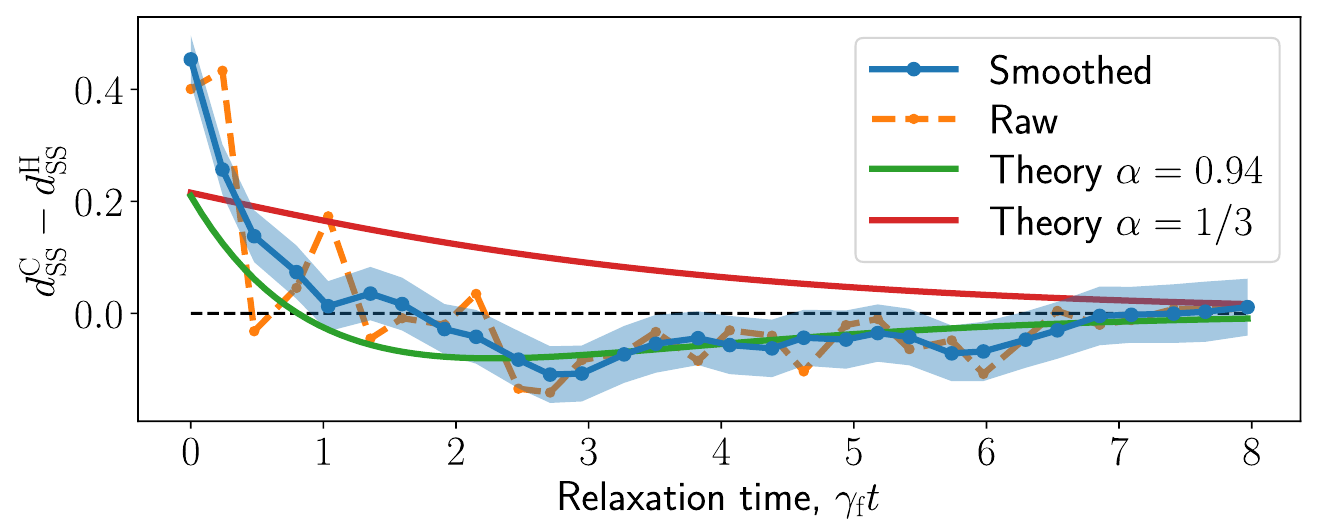}   
    \caption{Comparison between raw and smoothed data (orange and blue, respectively) and corresponding analytical predictions of the difference between distances from final state, $d_\text{ss}^\text{C}-d_\text{ss}^\text{H}$. The inverse-\ac{ME} seen as a negative value of the smoothed data, beyond the error bars, agrees with our prediction for $\alpha = 0.94$ (green). The coherence requirement is exemplified by presenting an alternative prediction, with $\alpha = \nicefrac{1}{3}$ (red), below the minimal value for a strong-\ac{ME}, which indeed shows no effect.}
    \label{fig:Crossing}
\end{figure}

Figure~\ref{fig:Crossing} presents the distance between these two systems during relaxation, $d_\text{ss}^\text{C}\leri{t}-d_\text{ss}^{H}\leri{t}$, for the raw data (orange) and smoothed data (blue). Indeed, $d_\text{ss}^\text{C}\leri{0}-d_\text{ss}^\text{H}\leri{0}>0$, indicating the cold system is initially at a larger distance from steady state. However, during the relaxation we observe a crossing time $t_\text{cross}$ after which $d_\text{ss}^\text{C}\leri{t}-d_\text{ss}^\text{H}\leri{t}<0$, beyond $\pm2\sigma$ error bars due to quantum shot-noise. The theoretical prediction for this distance at $\alpha=0.94$ is shown (green), with a well correspondence to the data. Furthermore we show the theoretical prediction for the case, $\alpha=1/3$, outside of the strong-\ac{ME} regime, in which no crossing is observed (red).

\begin{figure}[t]
    \includegraphics[width=1\textwidth]{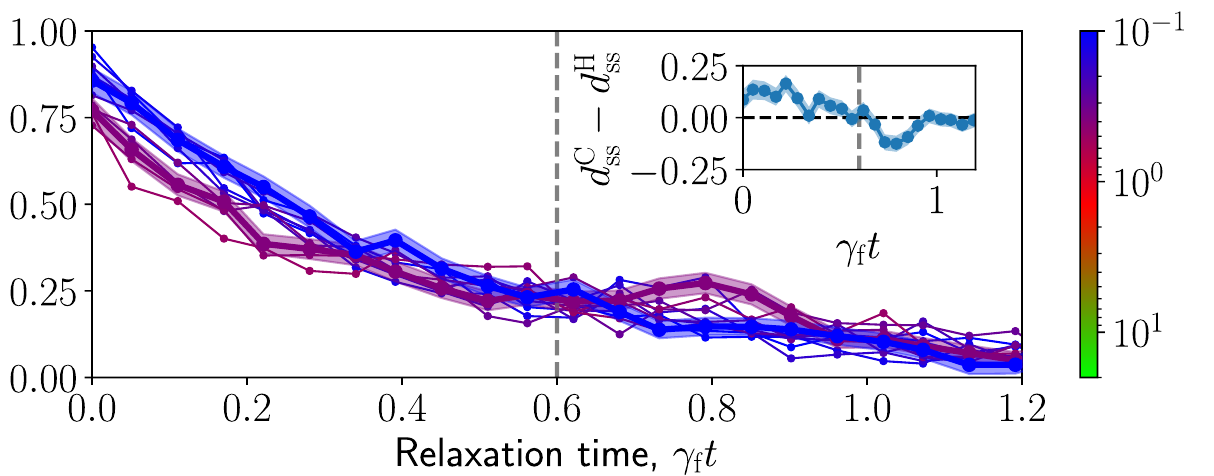}   
    \caption{The inverse-\ac{ME} realized with initial state preperation. The qubit's distance to the final steady state, $d_\text{ss}\leri{t}$, is shown for qubits initialized at different temperatures (color) and quenched to $\gamma_\text{f}'=100$ and $\alpha=0.94$. We highlight $d_\text{ss}^\text{C}$ (thick dark blue) and $d_\text{ss}^\text{H}$ (thick light blue), initially cold and hot systems, respectively, and present their $\pm2\sigma$ confidence intervals (shaded regions). As shown, the hot qubit starts closer to the final steady state. However, at $t_\text{cross}\approx0.6\gamma_f^{-1}$ (vertical grey) the cold system surpasses the hot system and relaxes first, manifesting the inverse Mpemba effect. Error bars (shaded regions) correspond to $\pm 2\sigma$ confidence intervals due to the quantum shot-noise of 400 experimental repetitions. Inset: $d_\text{ss}^\text{C}\leri{t}-d_\text{ss}^\text{H}\leri{t}$ highlighting the crossing of the two systems, beyond the confidence intervals. }
    \label{fig:DynamicsSP}
\end{figure}

Next, we directly prepare the qubit at an initial steady state. We write the initial steady state density matrix, $\rho_\text{i}^\text{ss}$, as a linear combination of two pure-states. Here the steady states are all of the form, $\rho_\text{i}^\text{ss}=\frac{1-p}{2}\ketbra{+\theta}{+\theta}+\frac{1+p}{2}\ketbra{-\theta}{-\theta}$, with $\theta$ and $p$ representing the direction and distance of the steady state from the Bloch sphere origin and $\ket{\pm\theta}\equiv\exp\left[-\frac{i}{2}\leri{\theta\pm\frac{\pi}{2}}\sigma_x\right]\ket{\uparrow}$~\cite{SM}. Then, the evolution of $\rho_\text{i}^\text{ss}$ is equivalent to the same linear combination of evolved pure states. Observables stemming from the evolution of $\rho_\text{i}^\text{ss}$, are recovered by measuring the same observables on the evolution of $\ket{\pm\theta}$, and using a weighted linear combination of the results, with weights $\leri{1\mp p}/2$.

Figure~\ref{fig:DynamicsSP} shows the dynamics of system initialized at steady states with respect to various  $\gamma_\text{i}^\prime$'s, and tracks their evolution as a function of time (horizontal) towards a fixed $\gamma_\text{f}^\prime=100>\gamma_\text{i}^\prime$. Similarly to Fig.~\ref{fig:DynamicsDigital}, we show the distance to the final steady state, $d_\text{ss}$, and highlight an initially cold, $\gamma_\text{i}^{\prime\text{C}}\approx0$ (blue) and hot, $\gamma_\text{i}^{\prime\text{H}}=0.390$ (purple) systems. As above, $d_\text{ss}^\text{C}\leri{0}>d_\text{ss}^\text{H}\leri{0}$, indicating the cold system is initially at a larger distance from steady state, yet during the relaxation we observe a crossing time $t_\text{cross}$ after which $d_\text{ss}^\text{C}\leri{t}<d_\text{ss}^\text{W}\leri{t}$, beyond error bars. This is also reflected in the inset which shows $d_\text{ss}^\text{C}-d_\text{ss}^\text{H}$ (vertical) initially positive and at later times negative, beyond the error bars.

In conclusion, we have proposed and experimentally demonstrated the inverse \ac{ME} on as single qubit. Furthermore, we have proven that a strong, i.e. exponentially faster relaxation, \ac{ME} exists only for a sufficiently coherent qubit. As our findings pertain the simplest quantum system, one expects to find the \ac{ME} in larger quantum systems, such as quantum computers, in which maintaining a low temperature for long times is crucial.
\begin{acknowledgments}
This work was supported by the Israel Science Foundation Quantum Science and Technology (Grant 3457/21).
O.R. is supported by the ISF (Grant 232/23).
G. T. is supported by the Center for Statistical Mechanics at the Weizmann Institute of Science, the Simons Foundation (grant 662962), the grants HALT and Hydrotronics of the EU Horizon 2020 program and the NSF-BSF (grant 2020765).
\end{acknowledgments}

\bibliography{references}

\end{document}

\title{The inverse Mpemba effect demonstrated on a single trapped ion qubit - Supplemental Material}
\author{Shahaf Aharony Shapira$^{\dagger}$}
\email{shahaf.aharony@weizmann.ac.il}
\author{Yotam Shapira$^{\dagger}$} 
\author{Jovan Markov}
\author{Gianluca Teza}
\author{Nitzan Akerman}
\author{Oren Raz}
\author{Roee Ozeri}
\affiliation{Department of Physics of Complex Systems\\
		Weizmann Institute of Science, Rehovot 7610001, Israel\\ 
		$^\dagger$ These authors contributed equally to this work
}

\maketitle

\onecolumngrid

\section{1. Analytic solution of the Lindblad equation} \label{sup:analyticSol}
%---------------------------
\begin{acronym}
	\acro{ME}{Mpemba effect}
	\acro{GKSL}{Gorini-Kossakowski-Sudarshan-Lindblad}
	\acro{dof}{degree of freedom}
	\acro{ODE}{ordinary differential equation}
	\acro{QCs}{quantum computers}
	\acro{SM}{Supplemental Material}
\end{acronym}
%---------------------------

We provide additional information concerning the solution of the qubit's dynamics. The Lindbladian in the main text can be transformed, using the Kronecker product, into a linear form: $\partial_t \vec{\rho} = \hat{L} \vec{\rho}$, where $\vec{\rho}$ is a column vector and $\hat{L}$ is a $4\times4$ matrix. For our model, given in~Eq. (1) of the main text, we obtain 
\begin{align}
	\hat{L} \leri{\Omega,\gamma,\alpha} &= \begin{pmatrix}
		-2 \alpha \gamma & -\frac{i \Omega}{2} & \frac{i \Omega}{2} & 0 \\
		-\frac{i \Omega}{2} & -\gamma &  0 & \frac{i \Omega}{2} \\
		\frac{i \Omega}{2} & 0 &  -\gamma  &  -\frac{i \Omega}{2} \\
		2 \alpha \gamma & \frac{i \Omega}{2} & - \frac{i \Omega}{2}  & 0 
	\end{pmatrix}   ,	 
\end{align}
such that the equation of motion reads,
\begin{align}
	\left\{ \begin{array}{l}
		\dot{\rho}_{11} = -2\alpha \gamma \rho_{11} + \frac{i \Omega}{2} \leri{\rho_{12} - \rho_{21} } \,, \\
		\dot{\rho}_{21} = \frac{i \Omega}{2} \leri{\rho_{22} - \rho_{11} } -\gamma \rho_{21} \,, \\
		\dot{\rho}_{12} = \frac{i \Omega}{2} \leri{\rho_{11} - \rho_{22} } - \gamma \rho_{12} \,, \\
		\dot{\rho}_{22} = 2\alpha \gamma \rho_{11} + \frac{i \Omega}{2} \leri{\rho_{21} - \rho_{12} } \,.
	\end{array} \right. 
\end{align}
Since $\rho=\rho^\dagger$ and $\text{Tr}\left[\rho\right]=1$, the state of a single qubit can be equivalently written in the Bloch vector representation:
\begin{align}
	\vec{r} &= \leri{ \rho_{12} + \rho_{21} ,\ i\leri{\rho_{12} - \rho_{21}} ,\ \rho_{11} - \rho_{22}} \equiv \leri{x, y, z}\,.
\end{align}
Thus, the set of ordinary differential equations is
\begin{align}
	\left\{ \begin{array}{l}
		\dot{x} = -\gamma x \,, \\
		\dot{y} = -\gamma y - \Omega z \,, \\
		\dot{z} =  - 2 \alpha \gamma \leri{1+z}  +\Omega y \,.
	\end{array} \right. 
\end{align}
The steady state solution of this system is given by:
\begin{align}
	\label{eqn:equiLocusSS}
	x^*&=0\,, &
	y^* &= \frac{\nicefrac{\Omega}{\gamma} }{1 + \frac{\Omega^2}{2 \alpha \gamma^2}} \,, &
	z^* &= -  \frac{1}{1+\frac{\Omega^2}{2 \alpha \gamma^2}}\,,
\end{align}
where the first coordinate, $x$, is decoupled from the dynamics with a stable fixed point (under the physical assumption $\gamma>0$). The remaining $2D$ first order differential equation for the $y-z$-coordinates is given by,
\begin{align} \label{eqn:ODEset}
	\dot{\vec{r}}_2 &= \begin{pmatrix}
		-\gamma & -\Omega \\
		\Omega & -2\alpha \gamma 
	\end{pmatrix} \vec{r}_2 + \begin{pmatrix}
		0 \\ -2\alpha \gamma
	\end{pmatrix} \,, &
	\vec{r}_2 &\equiv \begin{pmatrix}
		y \\ z
	\end{pmatrix}\,.
\end{align}
Their steady state solution are written in terms of the dimensionless variable $\gamma'\equiv \nicefrac{\gamma}{\Omega}$:
\begin{align}
	y^* &= \frac{1}{\gamma' \leri{1+\frac{1}{2 \alpha \gamma^{'2}}}} = - \frac{z^*}{\gamma'} \,, &
	z^* &= -  \frac{1}{1+\frac{1}{2 \alpha \gamma^{'2}}} = - \gamma' y^* \,.
\end{align}

These solutions, known as the steady state locus, form the right-half of an ellipse described by
\begin{align}
	\leri{\sqrt{\frac{2}{\alpha}}\ y^*}^2 + \leri{2z^*+1}^2 = 1 \,.\label{eqn:Elipse}
\end{align}
The $\theta$ and $p$ representation, utilized in the main text for direct state preparation, can be derived by setting, $y^\ast=p\cos\leri{\theta}$ and $z^\ast=p\sin\leri{\theta}$, yielding, $p=\frac{2\alpha\gamma^\prime\sqrt{1+\gamma^{\prime2}}}{1+2\alpha\gamma^{\prime2}}$ and $\theta=-\arctan\leri{\gamma^\prime}$.

\begin{figure}
	\begin{center}
		\includegraphics[width=0.5\textwidth]{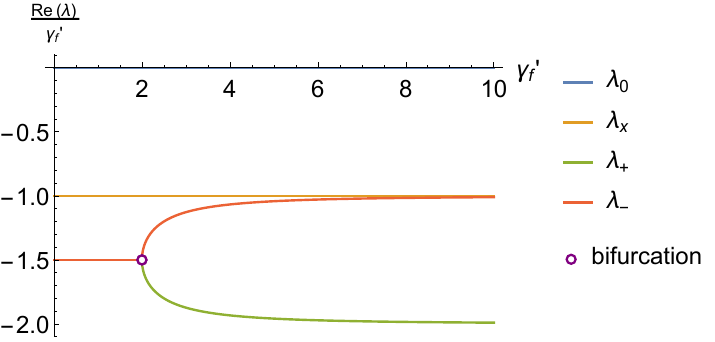}
	\end{center}	
	\caption{The real part of the system's eigenvalues over final temperature, for $\alpha=1$. The largest eigenvalue, corresponds to the steady state, zeros. For $H \propto \sigma_x $, the equation for $x$ decouples and the set can be solved analytically. In our case, $\lambda_x = -\gamma_\text{f}$. The last couple eigenvalues, $\lambda_\pm$, correspond to the dynamics in the $y-z$ plane and are therefore interesting. A \ac{ME} can only occur after the degeneracy is lifted.}  \label{plot:degeneracyPlot}
\end{figure}

The homogeneous part of Eq.~\eqref{eqn:ODEset} can be solved by diagonalization. The corresponding eigenvalues are given in Eq.~($4$) of the main text and are presented in Fig.~\ref{plot:degeneracyPlot}. Since the rates are dictated by their real parts, there exists a bifurcation point, $\gamma_b' \equiv \left| {\alpha - \nicefrac{1}{2}} \right|^{-1}$. The strong \ac{ME} exists only at final temperatures larger than the bifurcation point $\leri{ \gamma_\text{f} > \gamma_b}$, i.e. the point at which the degeneracy between eigenvalues breaks and there are distinct fast and slow relaxation modes.\\

The appropriate (normalized) eigenvectors are given by
\begin{align} \label{eq:eigenVectors}
	\vec{v}_{\pm} &= \frac{1}{2 \alpha -1} \begin{pmatrix}
		\frac{1}{\gamma_\text{f}^\prime \leri{\alpha - \nicefrac{1}{2} \pm \sqrt{\leri{\alpha - \nicefrac{1}{2}}^2 - \nicefrac{1}{\gamma_\text{f}^{'2}}}}} \\ 1
	\end{pmatrix} \,.
\end{align}

The coefficients of these modes are given by $a_n \leri{\gamma_\text{i}, \gamma_\text{f}} \equiv \vec{u}_n(\gamma_\text{f}) \cdot \vec{r}^\text{ ss}(\gamma_\text{i})$, where $\vec{u}_n$ are left eigenvectors, i.e row vectors satisfying the equation, 
\begin{equation}
	\vec{u}_n\begin{pmatrix}
		-\gamma & -\Omega \\
		\Omega & -2\alpha \gamma 
	\end{pmatrix}=\lambda_n \vec{u}_n.
\end{equation}
Explicitly, we obtain
\begin{align}
	a_\pm &= \frac{\alpha \cdot \leri{\alpha - \nicefrac{1}{2}} }{\sqrt{\leri{\alpha - \nicefrac{1}{2}}^2 - \nicefrac{1}{\gamma_\text{f}^{'2}}} \leri{2 \alpha + \nicefrac{1}{\gamma_\text{f}^{'2}}} \leri{2 \alpha + \nicefrac{1}{\gamma_\text{i}^{'2}}}} \leri{\frac{\gamma_\text{f}^{'}}{\gamma_\text{i}^{'}} -1} \times  \nonumber \\
	&  \left[ 4 \alpha \leri{\mp \alpha \pm \nicefrac{1}{2} + \sqrt{\leri{\alpha - \nicefrac{1}{2}}^2 - \nicefrac{1}{\gamma_\text{f}^{'2}}}} - \frac{2}{\gamma_\text{i}^{'} \gamma_\text{f}^{'}} \leri{\mp \alpha \mp \nicefrac{1}{2} + \sqrt{\leri{\alpha - \nicefrac{1}{2}}^2 - \nicefrac{1}{\gamma_\text{f}^{'2}}}} \pm \frac{2}{\gamma_\text{f}^{'2}} \right] 
	\,. \label{eq:cPlusMinus}
\end{align}
By inserting $\gamma_\text{i,SME}$ given in the main text, $a_-$ indeed vanishes.

\section{2. The absence of a strong direct-ME}
For a strong \ac{ME} to exist, the coefficient of slow relaxation $a_-$ must vanish. Therefore, in the case of a strong effect the dynamics is along the fast relaxation eigenvector, $\vec{v}_+$, only. The fast relaxation eigenvector of a given final temperature $\gamma_\text{f}'$ must therefore intersect the steady state locus in an additional point, corresponding to the ideal initial temperature $\gamma_\text{i, SME}'$.

\begin{figure}[h]
	\begin{center}
		\includegraphics[width=0.45\textwidth]{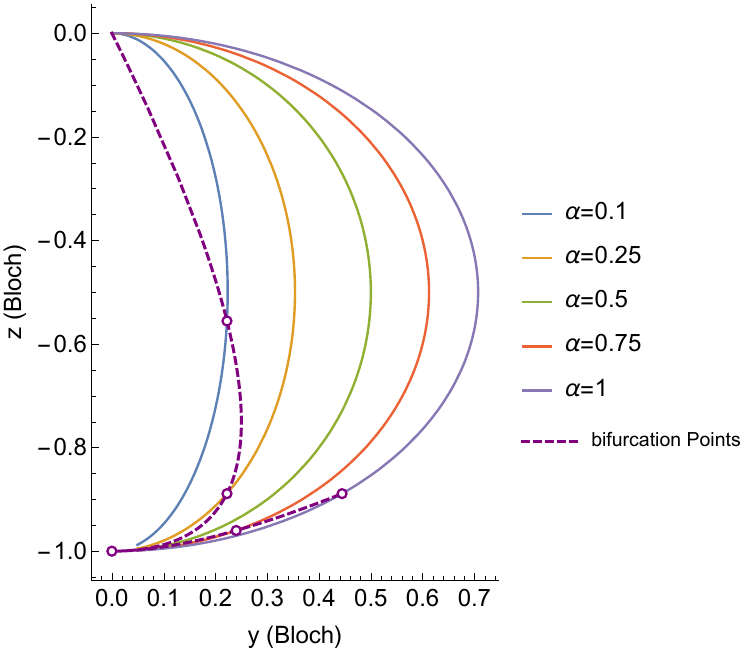}
		\caption{The steady state loci in the $y-z$ plane of the Bloch sphere, as given by Eq.~\eqref{eqn:equiLocusSS}, for different values of $\alpha$, as a function of temperature $\gamma' \in (0,20]$. The \ac{ME} exists only at temperatures above $\gamma_\text{b}$, represented by circles at the crossing with the bifurcation (purple dashed) line. We observe two bifurcation branches, however, only the lower one, crosses loci of $\alpha\geq\nicefrac{1}{2}$, is valid for a strong \ac{ME}.}
		\label{fig:equiLocusAlpha}
	\end{center}	
\end{figure}

For $\alpha>\frac{1}{2}$, both entries of $v_+$ are positive (See \eqref{eq:eigenVectors}), such that $v_+$ points to the up-right direction. On the steady state locus, the temperature is monotonic in $z$ such that states located higher on the locus necessarily correspond to lower temperatures. Moreover, since $\gamma_\text{f}' > \gamma_\text{b}'$ for a ME to occur, and the crossing of the steady state locus with the bifurcation-line of $\alpha> \frac{1}{2}$ is located in the southern-hemisphere of the locus (See Fig.~\ref{fig:equiLocusAlpha}), states located to the right on the locus also correspond to lower temperatures. The implications are that all possible fast vectors point to lower initial temperatures, relative to $\gamma_\text{f}'$. Accordingly, this system doesn't exhibit a strong, direct, \ac{ME}.

\begin{figure}[h]
	\begin{center}
		\includegraphics[width=0.6\textwidth]{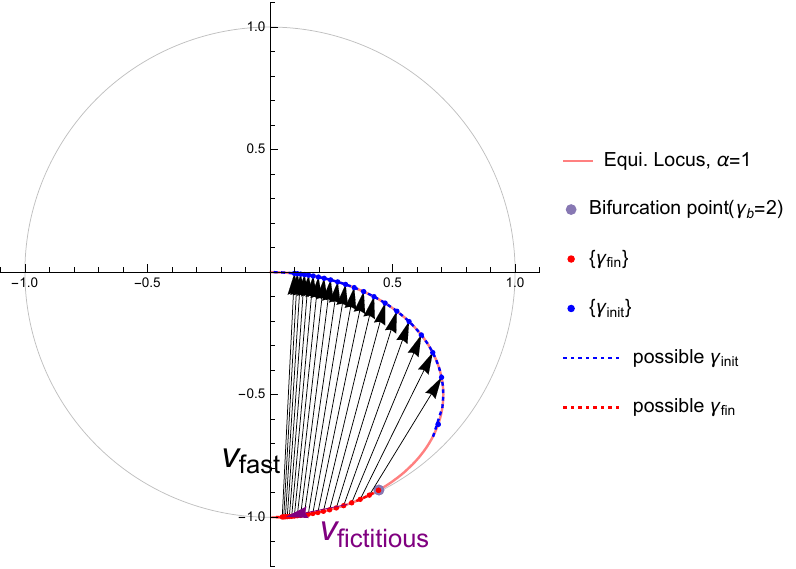}
	\end{center}	
	\caption{The steady state locus for $\alpha=1$ (pink line), inside the Bloch sphere (grey line), and three kinds of key points. The first is the bifurcation point, $\gamma_\text{b}'$ (purple dot), which is always located in the southern-hemisphere of the locus. Since the strong \ac{ME} exists only at final temperatures above this point, the second kind is $\gamma_\text{f}' > \gamma_\text{b}'$ (red dotted line). We show the fast relaxation vector, $v_+$, for various possible final temperatures (black lines). Each of these vectors intersects with the steady state locus at a point corresponds to $\gamma_\text{i, SME}'$, which is the initial temperature for the strong effect. Blue dotted line: the appropriate possible initial temperatures, for $\gamma_f > \gamma_b$. All possible initial temperatures are smaller then the bifurcation point, i.e. no strong direct effect. In other words, the absence of a vector pointing to higher temperatures, e.g. purple line, refutes the existence of a strong direct effect.}  \label{plot:fastVectorPlot}
\end{figure}

An explicit illustration of the orientation of $\vec{v}_+$, with respect to the steady state locus of $\alpha=1$, is presented in Fig.~\ref{plot:fastVectorPlot}.

\section{3. The existence of a strong inverse-ME}
\begin{figure}
	\begin{center}
		\includegraphics[width=0.45\textwidth]{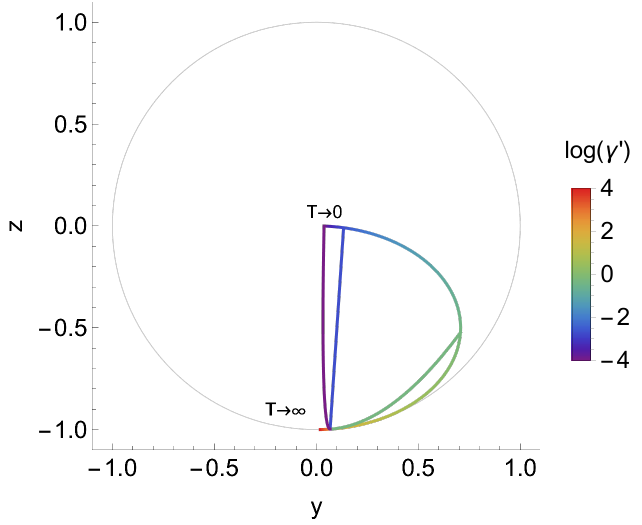}
		\hfill
		\includegraphics[width=0.45\textwidth]{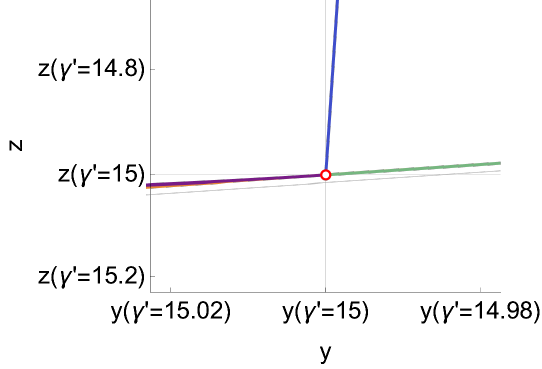}
		\caption{Left: The dynamics of three initial states towards the same final steady-state $\gamma_\text{f}'=15$ (red dot). In order to achieve a strong~\ac{ME} for that final temperature, the corresponding initial temperature is $\gamma_\text{SME,i}'\approx0.07$. Since at $\gamma_\text{i}^{\text{cold}\prime} \approx 0.07$ (blue line) the coefficient of slow relaxation vanishes, it exhibits an exponentially faster relaxation, compared, for example, to a colder $\gamma_\text{i}^{\text{colder}\prime}=0.02$ (purple line) and hotter $\gamma_\text{i}^{\text{hot} \prime} \approx 0.74$ (green line) initial states. Right: The evolution of $\gamma_\text{i}^\text{cold}$ is along the direction of $v_+$ only, and a zoom-in on the end of the process reveals how lower and higher initial temperatures approach relaxation from opposite directions, along $v_-$.}
		\label{fig:locusAndDynamics}
	\end{center}	
\end{figure}

In the main text we prove the existence of a strong inverse-\ac{ME} by a vanishing of the coefficient of the slow relaxation mode, $a_-$. As complementary proof to the existence of the strong \ac{ME}, one simply tracks the evolution of states, near the final steady state. In the long-time limit, fast relaxation modes have already decayed. Assuming $a_-$ crosses zero continuously - it changes signs. Thus, states initialized before and after the zero-crossing will approach the final steady state from opposite directions of $\vec{v}_-$ in the $y-z$ plane. An example is shown in Fig.~\ref{fig:locusAndDynamics}.\\

An experimental observation of the \ac{ME} involves comparing the dynamics of a cold and hot initial states, $d_\text{SS}^\text{C}\leri{t}$ and $d_\text{SS}^\text{H}\leri{t}$, respectively. Specifically, the inverse \ac{ME} is demonstrated by showing that the initially colder system reaches the final steady state faster, although it starts further away. Namely, at the beginning of the experiment $d_\text{SS}^\text{C}\leri{t=0}>d_\text{SS}^\text{H}\leri{t=0}$, whereas after some later time, $t_\text{cross}$, the colder system surpasses the hotter such that the distances are reversed, i.e. $d_\text{SS}^\text{C}\leri{t>t_\text{cross}}<d_\text{SS}^\text{H}\leri{t>t_\text{cross}}$ .

\begin{figure}[h]
	\begin{center}
		\includegraphics[width=0.4\textwidth]{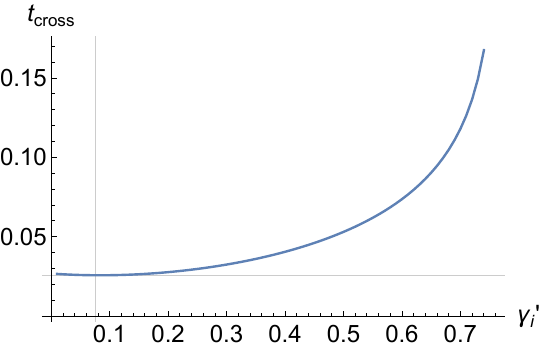}
		\hfill
		\includegraphics[width=0.4\textwidth]{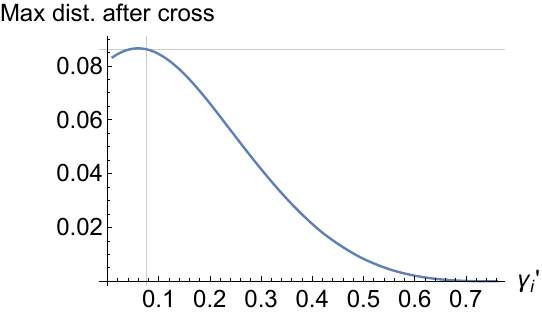}
	\end{center}	
	\caption{Left: The crossing time of states initialized at $\gamma_\text{i}^{\prime}$ compared to $\gamma_\text{i,h}^{\prime} = 0.77$, while relaxing to $\gamma_\text{f}'=15$. 
		Right: The corresponding maximal distance, post-crossing. The peak indicates the initial state yielding the most pronounced effect, $\gamma_\text{i,SME}^{\prime}$ (grey line), for which the effect is strong. }  \label{fig:CrossDmax}
\end{figure}

Measuring $d_\text{SS}^\text{C}\leri{t}<d_\text{SS}^\text{H}\leri{t}$ is inherently challenging, as the two quantities are exponentially decaying. The existence of a strong-\ac{ME} is helpful due to two reasons: Early onset of the crossing between the cold and hot distances from final steady state, and maximal distance post-crossing. The former allows for short coherence-time of the experimental apparatus and the latter requires less experimental repetitions. Both parameters are optimized at $\gamma_\text{i}\sim\gamma_\text{i,SME}$. Figure~\ref{fig:CrossDmax} shows this by fixing $\gamma_{i,H}^\prime=0.77$ and plotting the crossing time (left), as well as the maximal distance to various choices of $\gamma_{i,C}^\prime$ after the two distances cross (right). Clearly, the minimal crossing time, as well as maximal distance between the two decaying signals appear at $\gamma_\text{i}^\prime=\gamma_\text{i,SME}^\prime$ (vertical gray).

\begin{comment}
	\begin{figure}[h]
		\begin{center}
			\includegraphics[width=0.7\textwidth]{strongMEPlot.pdf}
			\includegraphics[width=0.5\textwidth]{strongMEPlotZoom.pdf}
		\end{center}	
		\caption{The values of initial temperature for which $a_-$ vanishes as a function of $\alpha$, assuming $\gamma_\text{f}'=100$. For $\gamma_i'$ to be real, $\alpha < \frac{1}{2} - \frac{1}{\gamma_\text{f}'}$ or $\alpha > \frac{1}{2} +  \frac{1}{\gamma_\text{f}'} $ (gray lines). Equivalently, $\gamma_f > \gamma_b$. Effectively, since $\gamma$ must also be positive, as it represents temperature, the strong \ac{ME} exists only for $\alpha > \frac{1}{2} +\frac{1}{\gamma_\text{f}'} $.}  \label{plot:strongMEPlot}
	\end{figure}
\end{comment} 

\section{4. Error analysis}

All error bars and error regions in the main text represent $\pm2\sigma$ confidence intervals due to quantum shot-noise, i.e. assuming a binomial distribution between two measurement outcomes. The corresponding experimental repetitions appearing in the main text.

These errors are propagated using standard methods, i.e. $\delta f\left(\boldsymbol{\sigma}\right)=\sqrt{\sum_i \left(\partial_{\sigma_i} f \cdot\delta \sigma_i\right)^2}$, with $\delta \sigma_i$ the error in the measurement of $\sigma_i\in\left\{\sigma_x,\sigma_y,\sigma_z\right\}$. This basic formula is used for calculating error bars of qubit tomography (e.g. in Fig. 2) as well as error bars of post-processed smoothed data (e.g. in Fig. 4).